\documentstyle[11pt,newpasp,twoside,epsf]{article}
\def\edcomment#1{\iffalse\marginpar{\raggedright\sl#1\/}\else\relax\fi}
\reversemarginpar

\begin{document}
\title{Variability in LINERs}
\author{J. C. Shields$^1$, H.-W. Rix$^2$, D. H. McIntosh$^3$, L. C. Ho$^4$,
G. Rudnick$^3$, A. V. Filippenko$^5$, W. L. W. Sargent$^6$, M. Sarzi$^{2,7}$,
and M. Eracleous$^8$}
\affil{$^1$Ohio University, Physics \& Astronomy Dept, Athens, OH 45701}
\affil{$^2$Max-Planck-Institut f\"ur Astronomie, Heidelberg, Germany}
\affil{$^3$Steward Observatory, University of Arizona, Tucson, AZ 85721}
\affil{$^4$Obs. of the Carnegie Institution of Washington, Pasadena, CA 91101}
\affil{$^5$University of California, Astronomy Dept, Berkeley, CA 94720-3411}
\affil{$^6$Palomar Observatory, Caltech 105-24, Pasadena, CA 91125}
\affil{$^7$Universit\'a di Padova, Dpto. di Astronomina, Padova, Italy}
\affil{$^8$Pennsylvania State University, Astronomy \& Astrophysics,  \\
University Park, PA 16802}

\begin{abstract}
A small number of LINERs have been seen to display variable H$\alpha$
emission with a very broad, double-peaked profile.  Recent
observations with the {\sl Hubble Space Telescope} indicate that such
emission may be a common attribute of LINERs.  The double-peaked or
double-shouldered line profiles resemble those found in a subset of
broad-line radio galaxies.  Several lines of argument suggest that
such features trace an outer thin accretion disk irradiated by an
inner ion torus, in accord with advection-dominated accretion flow
(ADAF) models.  Variability monitoring of this broad H$\alpha$
component thus may provide a means of testing accretion physics on
small scales within these sources.

\end{abstract}

\section{Introduction}
Low Ionization Nuclear Emission-line Regions (LINERs; Heckman 1980)
are found in approximately 30\% of bright galaxies (Ho, Filippenko, \&
Sargent 1997a).  While LINERs may form a heterogeneous class in terms
of excitation mechanisms (e.g., Filippenko 1996), an important
fraction are clearly weak manifestations of quasar-like phenomena, as
demonstrated by the presence of broad H$\alpha$ emission in $\ga 20$\%
of these sources (Ho et al. 1997b).

LINERs also vary in their luminous output, although rather little is
known about this aspect of their behavior.  Continuum variability has
been reported at radio (e.g., Heeschen \& Puschell 1983; Ho et
al. 1999b) and X-ray (e.g., Ho et al. 1999a) wavelengths, but the data
are limited to individual objects or small samples, observed with very
limited temporal coverage.  Measurement of variability in the optical
continuum is severely hampered by the weakness of the AGN continuum
relative to the starlight from the surrounding bulge.  Broad H$\alpha$
emission, when present, is also visible primarily as wings on the
H$\alpha$ + [N~{\sc ii}] narrow-line blend, requiring a careful
decomposition and continuum subtraction to measure accurately.

Monitoring of broad H$\alpha$ emission has been carried out for a few
LINERs, and suggests that variability in this feature is often weak or
absent (e.g., Ho, Filippenko, \& Sargent 1996).  However, variability
of a dramatic nature, involving the appearance of a broad,
double-peaked line component, has been reported in three objects:
NGC~1097 (Storchi-Bergmann, Baldwin, \& Wilson 1993), Pictor~A
(Halpern \& Eracleous 1994; Sulentic et al. 1995), and M81 (Bower et
al. 1996).  If this behavior is characteristic of LINERs, the nature
of this emission component and its temporal evolution may hold
important clues to the underlying accretion structure.

\section{New Results from HST}

New results that bear on accretion processes in LINERs have recently
emerged from a Survey of Nearby Nuclei with STIS (SUNNS; Rix et
al. 2000).  Under this program, optical long-slit spectra of 24
nearby, weakly active galaxy nuclei were obtained in order to study
the gas kinematics and ionization properties on small scales
(resolution $\approx 0.1\arcsec$).  A surprise that emerged from this
work was the discovery of ultra-broad H$\alpha$ emission in NGC~4203
and NGC~4450 (Figure 1).  Both of these objects were classified from
ground-based spectra as LINER 1.9 sources (i.e., showing indications
of broad H$\alpha$; Ho et al. 1997b), but the {\sl Hubble Space
Telescope} ({\sl HST}) spectra reveal for the first time emission
components showing double peaks or shoulders, separated by $\sim 7000$
km s$^{-1}$ in velocity (Shields et al. 2000; Ho et al. 2000).  
A similar finding based on {\sl HST} observations of the LINER 
NGC~4579 has recently been reported by Barth et al. (2000).  These
emission components bear a strong resemblance to the variable
double-peaked emission reported in the past for some other LINERs 
(\S 1).

\begin{figure}
\plotfiddle{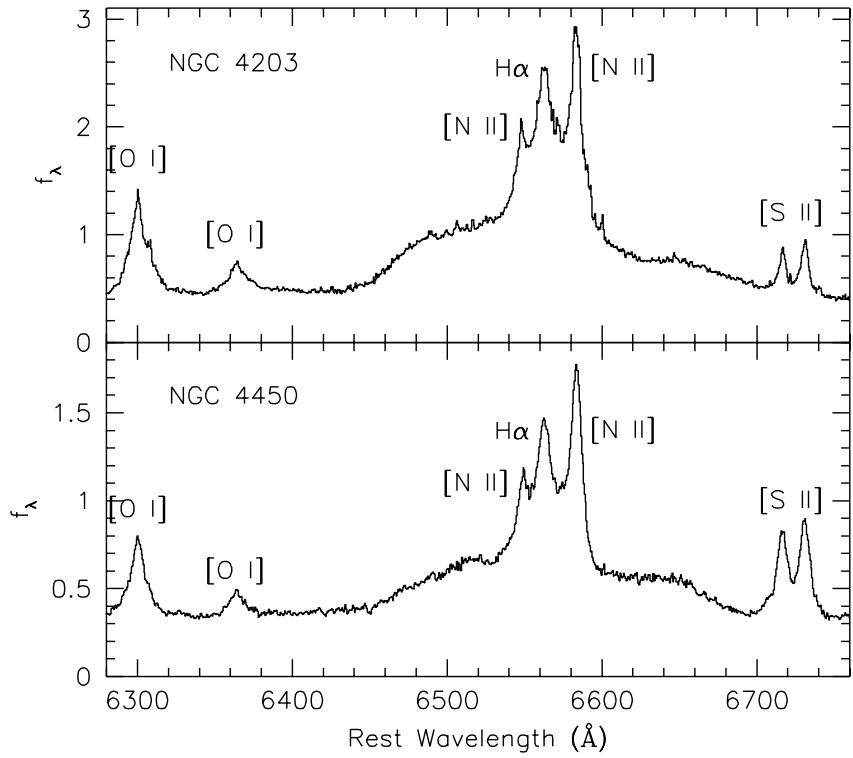}{4truein}{0}{80}{80}{-205}{-265}
\caption{STIS spectra of the central $0\farcs 25 \times 0\farcs 2$ 
for the LINERs NGC~4203 and NGC~4450.  Flux densities are in units
of $10^{-15}$ ergs s$^{-1}$ cm$^{-2}$ \AA$^{-1}$.}
\end{figure}
\begin{figure}
\plotone{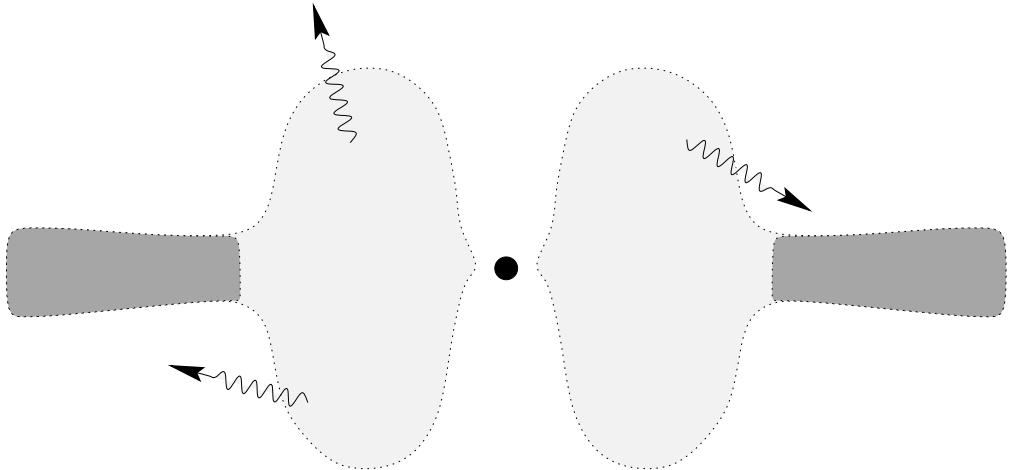}
\caption{Cartoon cross-section showing a black hole encircled by an
ion torus, with the latter irradiating an outer thin disk.  In this
picture the outer disk produces the observed double-peaked H$\alpha$
emission.}

\end{figure}

\section{Double-Peaked Emission}

Double-peaked H$\alpha$ emission similar to that found in LINERs has
been studied previously in other AGNs, notably in a subset of
broad-line radio galaxies (BLRGs; Eracleous \& Halpern 1994).
Double-peaked lines in general are suggestive of emission from a
rotating disk, and the H$\alpha$ profiles in BLRGs have been
successfully reproduced in a number of cases with relativistic accretion
disk models (e.g., Chen \& Halpern 1989; Eracleous \& Halpern 1994).
This interpretation is not unique, with alternative explanations
including emission from bipolar outflows, and from gas associated with
binary black holes (Eracleous 1999 and references therein).  However,
the accretion disk model remains attractive for several reasons,
including its ability to explain several characteristic asymmetries of
the line profiles that can naturally be ascribed to relativistic
effects.

Why would some accretion disks exhibit pronounced emission in the
Balmer lines, while others do not?  One explanation comes from
theoretical models of accretion flows described by small accretion
rates ($\dot M$), relative to the Eddington value (Rees et al. 1982;
Chen \& Halpern 1989).  In these scenarios, ions in the inner disk
cool inefficiently and become very hot; the resulting high pressure
produces a puffed-up structure, or ``ion torus,'' with nearly
spherical inflow.  Most of the thermal accretion energy of the torus
is carried into the black hole, rather than being emitted as
radiation, in contrast with the expected behavior of thin disks.  As a
source of energetic photons, however, the torus may nonetheless be
important for irradiating an outer disk that remains thin (Figure 2).
Reprocessing of this energy may then give rise to the double-peaked
emission.  Sources with higher $\dot M$ would retain a thin disk and
no torus, so that the added external illumination and double-peaked
emission would be absent.

This scenario fits well, at least in qualitative terms, with the
Advection Dominated Accretion Flow (ADAF) models discussed in detail
at this meeting by Narayan and Quataert.  The low luminosity of
LINERs is logically consistent with a low accretion rate and radiative
efficiency.  X-ray monitoring of LINERs also reveals weak variability
in these systems in comparison with Seyfert nuclei, a result in 
accord with ADAFs, for which the X-ray source is characteristically
larger than in thin-disk accretion sources (Ptak et al. 1998).

\section{Variability}

A natural question concerning the SUNNS discovery of double-peaked
LINER emission is whether this finding stems from variability, as in
the previous cases reported from the ground.  An alternative
explanation is that the double-shouldered emission was found because
of the substantial diminution of background starlight in the {\sl HST}
aperture in comparison with ground-based apertures.  The necessity of
removing the underlying stellar continuum presents a major difficulty
in the measurement of emission components for these weak AGNs, using
ground-based data.

The Palomar spectra of these sources published previously by Ho et
al. (1997b) unfortunately do not provide an unambiguous answer, since
wavelength-dependent focus variations render these data insensitive to
the detection of ultra-broad emission components.  However, the fact
that {\sl two} objects out of the (small) SUNNS sample exhibit this
property suggests that the aperture effect, rather than propitious
variability, is responsible.

While variability in NGC~4203 and NGC~4450 has not been demonstrated,
there are nonetheless good reasons for monitoring these sources in the
future in order to look for, and quantify, variability in the 
broad-line profile.  Changes in the line profile can be used to gain
information on characteristic time scales and dimensions for the
accretion structure.  In the LINER NGC~1097, the broad-line profile
showed a reversal over time in the relative height of the red and blue peaks,
indicating that asymmetries cannot be ascribed strictly to
relativistic effects, which would favor a higher blue peak for a
symmetric disk (Storchi-Bergmann et al. 1997).  Optical monitoring of
the nuclei of objects such as NGC~4203 and NGC~4450 will be
challenging, but would be feasible under conditions of good seeing,
with adaptive optics, or from space, such that background starlight is
minimized.  NGC~4203 is of particular interest in that the kinematics
of resolved nebular emission surrounding the nucleus provide an upper
limit on central black hole mass ($< 5 \times 10^6$ M$_\odot$), with
the promise of a measurement or more stringent limit from future
observations (Shields et al. 2000; Sarzi et al. 2000).

An easier task that is likely to be productive is to monitor the time
evolution of broad emission in a sample of BLRGs that show similar
line profiles in ground-based spectra.  Initial efforts of this type
have provided tantalizing indications of possible periodicities, and
changes in the line profile consistent with an orbiting structure or
wave in an accretion disk (Zheng, Veilleux, \& Grandi 1991; Newman et
al. 1997; Gilbert et al. 1999).  As discussed in the next section,
there are good reasons to think that conclusions drawn from monitoring
luminous BLRGs may apply also to LINERs.

\section{LINERs and BLRGs}

Some LINERs and BLRGs share the common trait of double-shouldered
H$\alpha$, but prototypes of each class contrast sharply in terms of
their total power and radio properties.  Heckman (1980) noted that
LINERs are often associated with radio emission, but these sources
tend to be weak cores (Nagar et al. 2000).  In comparison, BLRGs often
exhibit strong, extended jets and lobe emission, in addition to
powerful cores.  Does it make sense to view these disparate beasts as
a common phenomenon?

Several pieces of evidence suggest that BLRGs and LINERs are, in fact,
close cousins.  BLRGs showing double-peaked emission have relatively
strong low-ionization emission, in resemblance to LINERs (Eracleous \&
Halpern 1994).  But perhaps more fundamentally, LINERs, like the BLRGs,
are {\sl radio-loud} sources.  Observations with {\sl HST} make it
possible to separate out the AGN continuum from the background
starlight in LINERs.  Ho (1999) has recently used such data, in
conjunction with other multiwavelength measurements, to assemble the
broad-band spectral energy distributions (SEDs) for a sample of
LINERs.  The results, and radio/optical flux ratios, clearly
demonstrate that LINERs more closely resemble luminous radio-loud AGNs
than radio-quiet systems.  Since LINERs are the most common form of
active nucleus, and are mostly found in early-type disk galaxies (Ho
et al. 1997a), this statement inverts two pieces of canonical wisdom
concerning AGNs; i.e., when low-luminosity objects are included, {\sl
(1) the majority of AGNs are radio-loud}, and {\sl (2) most radio-loud
AGNs are found in disk galaxies}.

One notable aspect of LINER SEDs is that they are essentially missing
the big blue bump (BBB) that is typical of luminous AGNs.  (Note that
the classification of LINERs as radio-loud systems is supported by the
overall SED and not simply a relative weakness at optical
wavelengths.)  BLRGs with double-peaked emission lines also show
indications of relatively weak BBB emission (Eracleous \& Halpern
1994). The BBB is normally attributed to thermal emission from a
geometrically thin accretion disk.  The absence of this spectral
feature can be understood in the framework of the ADAF model, in that
the inner thin disk responsible for much of the short-wavelength
optical/UV emission in luminous systems is missing when the ADAF is
present (see Fig. 2).

\section{Conclusions}

The low luminosity of LINERs and their close association with the
large bulges of early-type disk galaxies makes the study of their
variability a challenging task in comparison with similar studies for 
Seyfert galaxies or quasars.  Yet results for a handful of LINERs
indicate that these sources do show variations in their broad lines,
that appear predominantly in a double-peaked component.  Results from
the SUNNS survey suggest that such double-peaked emission may be
a very common attribute of LINERs.  This emission resembles that of
more powerful BLRGs, and several lines of argument indicate that
LINERs and BLRGS are related phenomena.  Accretion via an ADAF
encircled by a residual thin disk provides a very appealing
phenomenological explanation for the properties of these objects,
in which case the broad H$\alpha$ emission acts as a direct
tracer of the accretion structure.  These results provide a strong
motivation for future variability monitoring of LINERs and BLRGs,
in order to probe the structure of these systems and test physical
models of the accretion process.

\acknowledgements

This work was supported financially by NASA grant NAG 5-3556, and by
GO-07361-96A, awarded by STScI, which is operated by AURA, Inc., for
NASA under contract NAS 5-26555.

\end{document}